\documentclass[12pt,a4paper]{article}

\usepackage{latexsym,amsmath}
\usepackage{amsfonts}
\allowdisplaybreaks[2]
\renewcommand{\title}[1]{\null\vspace{10mm}\noindent
                         {\Large{\bf #1}}\vspace{10mm}}
\newcommand{\authors}[1]{\noindent{\large #1}\vspace{20mm}}
\newcommand{\address}[1]{{\center{\noindent\small\itshape #1\vspace{0mm}}}}

\begin{document}
\begin{titlepage}

\begin{center}
\hspace*{\fill}{{\normalsize \begin{tabular}{l}
                              {\sf hep-th/0210288}\\
                              {\sf REF. TUW 02-022}
                              \end{tabular}  }}

\title{The Energy-Momentum Tensor in Noncommutative Gauge Field Models}

\authors{J.~M.~Grimstrup$^{1}$, B. Kloib\"ock$^{2}$,
 L.~Popp$^{3}$, V. Putz$^{4}$, M. Schweda$^{5}$,\\
 M. Wickenhauser$^{6}$}

\vspace{-15mm}

\address{$^{1,2,3,5,6}$  Institut f\"ur Theoretische Physik,
Technische Universit\"at Wien \\
Wiedner Hauptstra\ss e 8--10, A-1040 Wien, Austria}

\address{$^{4}$ Max-Planck-Institute for Mathematics in the Sciences \\
 Inselstr. 22 - 26, 04103 Leipzig, Germany}

\footnotetext[1]{jesper@hep.itp.tuwien.ac.at, work supported by
The Danish Research Agency.}

\footnotetext[3]{popp@hep.itp.tuwien.ac.at, work supported in part
by ``Fonds zur F\"orderung der Wissenschaftlichen Forschung'' (FWF)
under contract P13125-PHY.}

\footnotetext[4]{putz@mis.mpg.de, work supported 
 by ``Fonds zur F\"orderung der Wissenschaftlichen Forschung''
(FWF) under contract P13126-TPH}

\footnotetext[5]{mschweda@tph.tuwien.ac.at}

\footnotetext[6]{wick@hep.itp.tuwien.ac.at}
\vspace{10mm}

\begin{minipage}{12cm}
{\it Abstract.}
We discuss the different possibilities of constructing the various 
energy-momentum tensors for noncommutative gauge field models. We use
Jackiw's method in order to get symmetric and gauge invariant stress 
tensors---at least for commutative gauge field theories.
The noncommutative counterparts are analyzed with the same methods. The 
issues for the noncommutative cases are worked out.
  \vspace*{1cm}
\end{minipage}

\end{center}
\end{titlepage}

\section {Introduction}

In a previous paper \cite{Gerhold} we have analyzed the 
energy-momentum tensor on noncommutative spaces and we have found that
the dilation symmetry is broken due to the presence of the deformation 
parameter $\theta^{\mu\nu}$ characterizing the noncommutative geometry
by \cite{Filk}
 \begin{equation}
 [\hat x^\mu,\hat x^\nu] = i\theta^{\mu\nu}.
 \end{equation}
The existence of a constant, fixed antisymmetric tensor field 
$\theta^{\mu\nu}$ clearly also breaks the Lorentz symmetry \cite{Iorio},
\cite{Carroll}, \cite{Carlson} if 
$\theta^{\mu\nu}$ does not have a tensorial transformation behaviour with 
respect to Lorentz transformations. This situation resembles in
some sense the axial gauge in gauge field models. There, the 
presence of the constant, fixed gauge `direction' $n^\mu$ breaks the 
Lorentz invariance, too \cite{Boresch}. \\
In particular, the occurence of $\theta^{\mu\nu}$ in noncommutative 
quantum field models induces that the corresponding energy-momentum 
tensor needs neither be symmetric (for massless models), nor traceless.\\
The aim of the present work is the investigation of the construction of
the energy-momentum tensor in massless and commutative gauge field models
and their noncommutative counterparts, in order to work out the different 
aspects of the stress tensor for both cases.\\
Generally, the usual Noether procedure for the construction of the 
canonical energy-momentum tensor in the worst case needs an improvement
procedure and the Belinfante trick \cite{Callan}, \cite{Coleman}, 
\cite{Abou-Zeid} in order to get a symmetric and
traceless stress tensor. However, due to the idea of 
Jackiw \cite{Abou-Zeid}, \cite{Jackiw}, \cite{Bichl}, 
there is a more direct method to get the correct
stress tensor by combining the Noether procedure translations
with field dependent gauge transformations.\\
The paper is organized as follows. In section 2 we demonstrate the
power of Jackiw's recipe for the construction of the energy-momentum
tensor for a general $U(N)$ non-Abelian gauge field model. Both 
quantities, the canonical stress tensor and the symmetric one will be 
discussed in defining appropriate Ward-identity operators \cite{Kraus} for
the description of infinitesimal translations. Additionally, the
commutative Yang-Mills model is also characterized by the gauge symmetry
implying the existence of a conserved gauge current.\\
Section 3 is devoted to the discussion of the energy-momentum tensor for
noncommutative non-Abelian gauge field models. The methods used for the
commutative case can be applied without any difficulties also to the 
noncommutative counterpart. The results obtained in section 3 are very
similar to the corresponding output of the commutative case. However,
there is a severe difference: The stress tensors are no longer locally
conserved due to the fact that cyclic rotation of Moyal products occuring
in the stress tensors is locally impossible. After all, the symmetric 
version of the energy-momentum tensor is locally covariantly conserved with 
respect to the covariant derivative of the gauge symmetry.\\
In sections 4 and 5 we discuss the influence of the so called
Seiberg-Witten map on the construction of gauge invariant actions and 
the corresponding stress tensors. In section 4 we treat the 
noncommutative non-Abelian field model obtained via the Seiberg-Witten 
map to lowest order in $\theta^{\mu\nu}$ \cite{Seiberg}. 
Section 5 discusses the
special case of the $\theta$-deformed Maxwell theory ($U(1)$). The 
construction of the symmetric energy-momentum tensor confirms a 
recent result obtained by Kruglov \cite{Kruglov}.\\
All investigations are done in the classical approximation without
radiative corrections.    
  
\section{Energy-Momentum Tensor in Ordinary\\
Yang-Mills Theory}

In order to demonstrate the various possible constructions (canonical form,
Belinfante procedure, construction modulo a gauge transformation) of 
the energy-momentum tensor \cite{Gerhold}, \cite{Callan}, \cite{Coleman},
\cite{Abou-Zeid}, \cite{Jackiw}, let us start with a commutative 
Yang-Mills model, where the gauge field is matrix-valued, $A_\mu = 
A_\mu^a X^a$, $X^a$ being the corresponding generators of the gauge group
$U(N)$.\\
The corresponding infinitesimal gauge transformation is given by
 \begin{equation}
 \delta_\lambda A_\mu = \partial_\mu\lambda - i[A_\mu,\lambda]=:D_\mu\lambda,
 \label{gaugesym}
 \end{equation}
implying that the non-Abelian field strength 
 \begin{equation}
 F_{\mu\nu}= \partial_\mu A_\nu -\partial_\nu A_\mu - i[A_\mu,A_\nu]
 \end{equation}
transforms covariantly,
 \begin{equation}
 \delta_\lambda F_{\mu\nu}= i[\lambda, F_{\mu\nu}]. \label{gaugesym2}
 \end{equation}
Therefore, the gauge invariant non-Abelian action is given by
 \begin{equation}
 \Gamma_{inv}[A] =- {1\over 4}\int d^4\!x\, tr\big( F_{\mu\nu}F^{\mu\nu}\big)
 =: - {1\over 4}\int d^4\!x\, tr F^2
 \end{equation}
The equation of motion for the gauge field is
 \begin{eqnarray}
 {\delta\over \delta A_\nu} \Gamma_{inv}[A] = D_\rho F^{\rho\nu} = 0.
 \end{eqnarray}
The symmetry transformation (\ref{gaugesym}) may be expressed by the global
Ward-identity (WI)-operator
 \begin{equation}
 W^G(\lambda) =\int d^4\!x\, tr D_\mu \lambda(x) {\delta\over \delta A_\mu(x)}.
 \end{equation}
Gauge invariance is stated through $W^G(\lambda)\Gamma_{inv} = 0.$ 
This implies the following local identity
for the gauge symmetry,
 \begin{equation}
 {\delta\over\delta\lambda(x)}W^G(\lambda)\Gamma_{inv} =0. 
 \end{equation}
This defines the locally conserved current for the gauge symmetry,
 \begin{equation}
 j^\mu_G =-i[A_\rho,F^{\rho\mu}] ,\quad  -\partial_\mu j^\mu_G=i\partial_\mu
 [A_\rho,F^{\rho\mu}] = 0. \label{gaugecur}
 \end{equation}
By direct computation and by use of the equation of motion one easily
verifies (\ref{gaugecur}).\\
Now we want to discuss the infinitesimal translation described by the following
global WI-operator,
 \begin{equation}
 W_\mu^T = \int d^4\!x\, tr \partial_\mu A_\nu(x){\delta\over \delta A_\nu(x)}.
 \label{transwi}
 \end{equation}
By applying the WI-operator (\ref{transwi}) to the gauge invariant action
one gets the canonical energy-momentum tensor due to translational invariance,
 \begin{eqnarray}
 W_\mu^T \Gamma_{inv} &=& \int d^4\!x\, tr\big(\partial^\rho({1\over2}
 \{F_{\rho\nu},
 \partial_\mu A^\nu\}-{1\over4}g_{\rho\mu}F^2)\big)\nonumber \\
 &=&-\int d^4\!x\, \partial^\rho T^c_{\rho\mu} = 0.
 \end{eqnarray}
Thus, the canonical energy-momentum tensor is defined as
 \begin{equation}
 T^c_{\rho\mu} := -tr\big({1\over2}\{F_{\rho\nu},
 \partial_\mu A^\nu\}-{1\over4}g_{\rho\mu}F^2\big).
 \end{equation}
It is simple to show that $T^c_{\rho\mu}$ is locally conserved by using
the equation of motion.\\
However, $T^c_{\rho\mu}$ is not gauge invariant, not traceless and not 
symmetric in
$(\rho,\mu)$. In order to  obtain a symmetric stress tensor 
one has two possibilities \cite{Callan}, \cite{Coleman}.
Here we follow the method proposed originally by R. Jackiw \cite{Jackiw} 
in using an alternative representation for infinitesimal translations. 
Modulo a field dependent gauge transformation a possible description of 
translations is given by
 \begin{equation}
 W_\mu^F = \int d^4\!x\, tr F_{\mu\nu}(x){\delta\over\delta A_\nu(x)},
 \end{equation}
leading to
 \begin{eqnarray}
 W_\mu^F\Gamma_{inv}&=&\int d^4\!x\, tr\big(\partial^\rho({1\over2}\{F_{\rho\nu},
 F_\mu^{\ \nu}\}-{1\over4}g_{\rho\mu}F^2)\big)\nonumber \\
 &=&-\int d^4\!x\, \partial^\rho T^s_{\rho\mu} = 0,
 \end{eqnarray}
where $T^s_{\rho\mu}$ is gauge invariant, symmetric and traceless,
 \begin{equation}
 T^s_{\rho\mu} := -tr\big({1\over2}\{F_{\rho\nu},
 F_\mu^{\ \nu}\}-{1\over4}g_{\rho\mu}F^2\big).
 \end{equation}
One observes that the Jackiw construction unifies the Belinfante and 
improvement procedure.\\
Using the splitting
 \begin{equation}
 F_{\mu\nu}= \partial_\mu A_\nu -\partial_\nu A_\mu - i[A_\mu,A_\nu]
 =\partial_\mu A_\nu - D_\nu A_\mu,
 \end{equation}
one gets for the canonical tensor
 \begin{equation}
 T^c_{\rho\mu} = -tr\big({1\over2}\{F_{\rho\nu},
 F_\mu^{\ \nu}+D^\nu A_\mu\}-{1\over4}g_{\rho\mu}F^2\big),
 \end{equation}
implying that the difference between the canonical tensor and the symmetric one
becomes
 \begin{equation}
 T^c_{\rho\mu}- T^s_{\rho\mu} = -{1\over2}tr \{F_{\rho\nu},D^\nu A_\mu\}.
 \end{equation}
Due to the fact that the WI-operator of the translation is represented by
 \begin{equation}
 W^T_\mu = W^F_\mu + W^G_\mu = \int d^4\!x\, tr\big( F_{\mu\nu}(x){\delta\over
 \delta A_\nu(x)}+ D_\nu A_\mu(x){\delta\over\delta A_\nu(x)}\big),
 \end{equation}
the field dependent gauge transformation corresponds to the difference 
$T^c_{\rho\mu}- T^s_{\rho\mu}$,
 \begin{eqnarray} 
 -W^G_\mu\Gamma_{inv}&=&-\int d^4\!x\, tr\big(D_\nu A_\mu(x){\delta\Gamma_{inv}
 \over \delta A_\nu(x)}\big) \nonumber \\
 &=&-\int d^4\!x\,tr\partial_\rho\big({1\over2}\{F_{\rho\nu},D^\nu A_\mu\}\big).
 \end{eqnarray}
This is easily checked by explicit calculation with the use of partial
integration,
 \begin{eqnarray}
 -W^G_\mu \Gamma_{inv} &=& -\int d^4\!x\, tr D_\nu A_\mu D_\rho F^{\rho\nu}
 \nonumber\\
 &=& -\int d^4\!x\, tr\big(\partial^\rho({1\over2}\{F_{\rho\nu},D^\nu A_\mu\})
 -D_\rho D_\nu A_\mu F^{\rho\nu}\big).
 \end{eqnarray}
With the antisymmetry of $F^{\rho\nu}$ the second term is easily shown 
to vanish. This is very similar to the construction of the stress tensor
of the Maxwell theory \cite{Jackson}. 
Another interesting comment has to be made. If we omit the 
$tr$ symbol in the definition of the energy 
momentum tensor $T^s_{\rho\mu}$,
 \begin{equation}
 T^{s\ast}_{\rho\mu} := -\big({1\over2}\{F_{\rho\nu},
 F_\mu^{\ \nu}\}-{1\over4}g_{\rho\mu}F^2\big),
 \end{equation}
we get an object which is (due to the equation of motion and the 
Bianchi identity) $covariantly$ conserved,
 \begin{equation}
 D^\rho T^{s\ast}_{\rho\mu} =0.\label{covcons}
 \end{equation}
In discussing the noncommutative counterpart we will find that a similar
`covariant conservation' is also valid there. With (\ref{covcons})
one finds
 \begin{equation}
 \partial^\rho T^{s\ast}_{\rho\mu} =i[A^\rho,T^{s\ast}_{\rho\mu}],
 \end{equation}    
which is consistent with
 \begin{equation}
 W_\mu^F\Gamma_{inv} = -\int d^4\!x\, tr(\partial^\rho T^{s\ast}_{\rho\mu})
 = 0.
 \end{equation}

\section{Energy-Momentum Tensor in Noncommutative \\
Yang-Mills Theory}

It is now straigthforward to discuss also the noncommutative structure in the
spirit of the considerations done in the previous section. In  
noncommutative gauge field models one has to replace all field products 
by $\star$-products \cite{Filk} and one introduces the noncommutative matrix 
valued gauge field $\hat A$. The corresponding gauge invariant action is
therefore 
\begin{equation}
 \hat\Gamma_{inv}[\hat A] =- {1\over 4}\int d^4\!x\, tr\big(\hat F_{\mu\nu}
 \star\hat F^{\mu\nu}\big), \label{start}
 \end{equation}
with the noncommutative field strength
 \begin{equation}
 \hat F_{\mu\nu}= \partial_\mu \hat A_\nu -\partial_\nu \hat A_\mu - 
 i[\hat A_\mu,\hat A_\nu]_M . 
 \end{equation}
Here the Moyal commutator is given by
 \begin{equation}
 [\hat A_\mu,\hat A_\nu]_M := \hat A_\mu\star\hat A_\nu -\hat A_\nu\star\hat 
 A_\mu, \label{moycom}
 \end{equation}
using the $\star$-product,
 \begin{equation}
 A(x)\star B(x) = e^{{i\over2}\theta^{\mu\nu}\partial^\xi_\mu\partial^\eta_\nu}
 A(x+\xi)B(x+\eta)\arrowvert_{\xi=\eta=0}.
 \end{equation}
The infinitesimal gauge transformation is defined as
 \begin{equation}
 \delta_{\hat\lambda}\hat A_\mu = \partial_\mu\hat\lambda - 
 i[\hat A_\mu,\hat\lambda]_M=:\hat D_\mu\star\hat\lambda,\label{ncgaugesym}
 \end{equation}
where $\hat\lambda$ is the noncommutative counterpart of $\lambda$ of
equation (\ref{gaugesym})
The equation of motion for the gauge field is then
 \begin{eqnarray}
 {\delta\over \delta \hat A_\nu} \hat \Gamma_{inv}[\hat A] =
 \hat  D_\rho\star \hat F^{\rho\nu} = 0,
 \end{eqnarray}
and for the locally conserved gauge current we get
 \begin{equation}
 \hat j^\mu_G =-i [\hat A_\rho,\hat F^{\rho\mu}]_M.
 \end{equation}
At the level of noncommutative gauge field models one can perform the same
steps as in the previous section. With
 \begin{eqnarray}
 \hat W_\mu^T\hat\Gamma_{inv}&=&{1\over 2}\int d^4\!x\, tr
 \big(\partial_\mu \hat A_\nu(x) 
 \star{\delta \hat\Gamma_{inv}\over \delta \hat A_\nu(x)}+{\delta 
 \hat\Gamma_{inv}\over\delta\hat A_\nu(x)}\star\partial_\mu \hat A_\nu(x)\big)
 \nonumber\\ 
 &=&\int d^4\!x\, tr\big(\partial^\rho({1\over2}
 \{\hat F_{\rho\nu}, \partial_\mu \hat A^\nu\}_M-
 {1\over4}g_{\rho\mu}\hat F_{\alpha\beta}\star \hat 
 F^{\alpha\beta})\big)
 \end{eqnarray}
we find
 \begin{equation}
 \hat T^c_{\rho\mu} = -\big({1\over2}\{\hat F_{\rho\nu},
 \partial_\mu \hat A^\nu\}_M-{1\over4}g_{\rho\mu}\hat F_{\alpha\beta}
 \star\hat F^{\alpha\beta}\big). \label{curc}
 \end{equation}
Analogously we have
 \begin{eqnarray}
 \hat W_\mu^F\hat\Gamma_{inv}&=&{1\over 2}\int d^4\!x\, 
 tr\big(\hat F_{\mu\nu}(x)\star 
 {\delta \hat\Gamma_{inv}\over \delta \hat A_\nu(x)}+{\delta 
 \hat\Gamma_{inv}\over \delta \hat A_\nu(x)}\star\hat F_{\mu\nu}(x)\big)
 \nonumber \\
 &=&\int d^4\!x\, tr\big(\partial^\rho({1\over2}\{\hat F_{\rho\nu},
 \hat F_\mu^{\ \nu}\}_M-{1\over4}g_{\rho\mu}\hat F_{\alpha\beta}\star 
 \hat F^{\alpha\beta})\big)
\end{eqnarray}
and
 \begin{equation}
 \hat T^s_{\rho\mu} := -\big({1\over2}\{\hat F_{\rho\nu},
 \hat F_\mu^{\ \nu}\}_M-{1\over4}g_{\rho\mu}\hat F_{\alpha\beta}\star
 \hat F^{\alpha\beta}\big). \label{curs}
 \end{equation}
Here $\{\ ,\ \}_M$ represents the Moyal anti-commutator in the sense of 
(\ref{moycom}). Note that in order to define `local' quantities, 
$\int d^4\!x\, tr$---the integration over 
space-time and the trace over the colour indices---cannot be 
seperated in noncommutative geometry. After all,  
(\ref{curs}) is symmetric, traceless and transforms covariantly 
with respect to (\ref{ncgaugesym}),\cite{Abou-Zeid}.
 It is also locally $covariantly$ conserved,
 \begin{equation}
 \hat D^\rho\star \hat T^s_{\rho\mu} = 0.
 \end{equation}
This is shown with the help of the equation of motion and the noncommutative
analogon to the Bianchi-identity. We find that the energy-momentum tensors 
are not locally conserved\footnote{Even a `formal' definition of the tensors
including the $tr$-symbol does not help, since the trace of a Moyal
commutator is $not$ vanishing locally.},
 \begin{equation}
 \partial^\rho\hat T^s_{\rho\mu} \neq 0\neq \partial^\rho\hat T^c_{\rho\mu},
 \end{equation}
which is already known from the works \cite{Gerhold}, \cite{Kruglov}.

\section{Enery-Momentum Tensor \\ via Seiberg-Witten Map}

Another possibility to formulate noncommutative gauge field models 
is based on the fact that one can use the so called Seiberg-Witten (SW-)map
\cite{Seiberg}, \cite{Madore}, \cite{Bichl2}. This map ensures the gauge equivalence between an ordinary gauge 
field and its noncommutative counterpart. It implies that the noncommutative
gauge field $\hat A_\mu$ and also $\hat F_{\mu\nu}$ can be expanded in a
series in the deformation parameter $\theta^{\mu\nu}$ of the noncommutative
space-time geometry, with coefficients depending on the ordinary gauge field.
This section discusses the translation invariance of the 
SW-expansion of the noncommutative $U(N)$-Yang Mills (NCYM-)theory.
The starting point is equation (\ref{start}).  
 \begin{equation}
 \hat\Gamma_{inv}[\hat A]=-{1\over 4}\int d^4\!x\, 
 tr\big(\hat F_{\mu\nu}\star\hat 
 F^{\mu\nu}\big).
 \end{equation}
The SW-map to lowest order in $\theta^{\mu\nu}$ for the noncommutative 
gauge field is
 \begin{equation}
 \hat A_\mu = A_\mu - {1\over4}\theta^{\rho\sigma}\{A_\rho,\partial_\sigma
 A_\mu + F_{\sigma\mu}\},
 \end{equation}
implying the following field strength expansion \cite{Seiberg}
 \begin{equation}
 \hat F_{\mu\nu} = F_{\mu\nu} + {1\over 4}\theta^{\rho\sigma}\big(
 2\{F_{\mu\rho},F_{\nu\sigma}\}-\{A_\rho, D_\sigma F_{\mu\nu} +\partial_\sigma
 F_{\mu\nu}\}\big).
 \end{equation}
For the ordinary (commutative, Lie-algebra valued) field $A_\mu$ and 
field strength $F_{\mu\nu}$ the 
corresponding gauge transformations are given by (\ref{gaugesym}) 
and (\ref{gaugesym2}), respectively. \\
Expanding the $\star$-product in (\ref{start}) we have \cite{Bichl2}, 
\cite{Jurco}
 \begin{eqnarray}
 \Gamma^\theta_{inv}[A] &=&\int d^4\!x\, tr\big(- {1\over 4}F_{\mu\nu}F^{\mu\nu}
 -{1\over2}\theta^{\alpha\beta}(F_{\mu\alpha} F_{\nu\beta}F^{\mu\nu}
 -{1\over4}F_{\alpha\beta}F^2)\big) + {\cal O}(\theta^2)\nonumber \\
 &=& \int d^4\!x\, tr{\cal L}^\theta_{inv} + {\cal O}(\theta^2) .
 \end{eqnarray}
The corresponding equation of motion is
 \begin{eqnarray}
 -D_\rho\Pi^{\rho\nu} &=& D_\rho\big( F^{\rho\nu} -{1\over4}
 (\theta^{\rho\nu}F_{\alpha\beta}F^{\alpha\beta} +\{F^{\rho\nu},
 \theta^{\alpha\beta}F_{\alpha\beta}\})\nonumber \\
 &&+{1\over 2}(\{\theta^{\nu\beta}F_{\alpha\beta},F^{\rho\alpha}\}
 -\{\theta^{\rho\beta}F_{\alpha\beta},F^{\nu\alpha}\}
 +\{F^\rho_{\ \alpha}, \theta^{\alpha\beta}F^\nu_{\ \beta}\})\big)\nonumber\\
 &=& 0.
 \end{eqnarray}
The quantity $\Pi^{\rho\nu}$ is antisymmetric,
 \begin{equation}
 \Pi^{\rho\nu}=-\Pi^{\nu\rho} = {\partial {\cal L}^\theta_{inv}\over \partial
 (\partial_\rho A_\nu)},
 \end{equation}
and $\Pi^{\rho 0}$ is the canonical momentum. The analogous 
calculations as in section 1 give now
 \begin{eqnarray}
 W_\mu^T \Gamma^\theta_{inv} &:=&\int d^4\!x\, tr\partial_\mu
 A_\nu(x){\delta\Gamma^\theta_{inv}\over\delta A_\nu(x)}\nonumber \\ 
 &=& \int d^4\!x\, tr\big(\partial^\rho({1\over2}\{-\Pi_{\rho\nu},
 \partial_\mu A^\nu\}+g_{\rho\mu}{\cal L}^\theta_{inv})\big) \nonumber \\
 &=&-\int d^4\!x\, \partial^\rho T^{c,\theta}_{\rho\mu} = 0.
 \end{eqnarray}
Thus, the canonical energy-momentum tensor becomes
 \begin{equation}
 T^{c,\theta}_{\rho\mu} := tr\big({1\over2}\{\Pi_{\rho\nu},
 \partial_\mu A^\nu\}-g_{\rho\mu}{\cal L}^\theta_{inv}\big).\label{thetcurc}
 \end{equation}
Similarly, one gets
 \begin{eqnarray}
 W_\mu^F \Gamma^\theta_{inv} &=& \int d^4\!x\, tr F_{\mu\nu}(x)
 {\delta\Gamma^\theta_{inv}\over\delta A_\nu(x)}\nonumber \\
 &=& \int d^4\!x\, tr\big(\partial^\rho({1\over2}\{-\Pi_{\rho\nu},
 F_\mu^{\ \nu}\}+g_{\rho\mu}{\cal L}^\theta_{inv})\big) \nonumber \\
 &=&-\int d^4\!x\, \partial^\rho T^{s,\theta}_{\rho\mu} = 0,
 \end{eqnarray}
implying the following definition,
 \begin{equation}
 T^{s,\theta}_{\rho\mu} := tr\big({1\over2}\{\Pi_{\rho\nu},
 F_\mu^{\ \nu}\}-g_{\rho\mu}{\cal L}^\theta_{inv}\big).\label{thetcurs}
 \end{equation}
Both currents (\ref{thetcurc}) and (\ref{thetcurs}) are locally 
conserverd,
 \begin{equation}
 \partial^\rho T^{c,\theta}_{\rho\mu} = \partial^\rho T^{s,\theta}_{\rho\mu}
 =0,
 \end{equation}
and they are related by a Belinfante like procedure
 \begin{eqnarray}
 T^{s,\theta}_{\rho\mu}&=&T^{c,\theta}_{\rho\mu}+
 tr\big(D^\nu(A_\mu \Pi_{\rho\nu})\big)\nonumber \\
 &=&  T^{c,\theta}_{\rho\mu} + \partial^\nu \chi_{[\nu\rho]\mu}.
 \end{eqnarray}
One observes that both versions of the enery-momentum tensor are neither 
symmetric nor traceless. This is due to the fact that the Lorentz invariance  
and the dilation symmetry are no longer maintained \cite{Bichl}.
However, one has to stress that $T^{s,\theta}_{\rho\mu}$ is invariant with
respect to infinitesimal gauge transformations (\ref{gaugesym}).

\section{The $U(1)$-case: $\theta$-deformed Maxwell theory}

The most simple, but still interesting, case of a $\theta$-expanded gauge 
theory is the $U(1)$-NCYM-theory, the  $\theta$-deformed Maxwell theory (without 
sources). One just replaces in the expressions derived in the previous 
section the matrix-valued $U(N)$ gauge field $A^aX^a$ by the ordinary
photon field. Omitting the trace symbols we get
 \begin{eqnarray}
 \Gamma^\theta_{inv}[A] &=&\int d^4\!x\, \big(- {1\over 4}F_{\mu\nu}F^{\mu\nu}
 -{1\over2}\theta^{\alpha\beta}(F_{\mu\alpha} F_{\nu\beta}F^{\mu\nu}
 -{1\over4}F_{\alpha\beta}F^2)\big) + {\cal O}(\theta^2)\nonumber \\
 &=& \int d^4\!x\, {\cal L}^\theta_{inv} + {\cal O}(\theta^2) .
 \end{eqnarray}
Again, with the canonical momentum,    
\begin{eqnarray}
 \Pi^{\rho\nu}&=& {\partial {\cal L}^\theta_{inv}\over 
 \partial(\partial_\rho A_\nu)} = 
 - F^{\rho\nu} +{1\over4}
 (\theta^{\rho\nu}F_{\alpha\beta}F^{\alpha\beta}) + {1\over2} F^{\rho\nu}
 \theta^{\alpha\beta}F_{\alpha\beta}\nonumber \\
 &&-(\theta^{\nu\beta}F_{\alpha\beta}F^{\rho\alpha}
 -\theta^{\rho\beta}F_{\alpha\beta}F^{\nu\alpha})
 -F^\rho_{\ \alpha}\theta^{\alpha\beta}F^\nu_{\ \beta},
\end{eqnarray}
we find the equation of motion \cite{Kruglov},
 \begin{equation}
 \partial_\rho\Pi^{\rho\nu} =0.
 \end{equation}
The stress tensors read
 \begin{eqnarray}
 T^{c,\theta}_{\rho\mu} &=& \Pi_{\rho\nu}\partial_\mu A^\nu
 -g_{\rho\mu}{\cal L}^\theta_{inv}, \\
 T^{s,\theta}_{\rho\mu} &=& \Pi_{\rho\nu}F_\mu^{\ \nu}
 -g_{\rho\mu}{\cal L}^\theta_{inv}.
 \end{eqnarray}
Explicitly we have for $T^{s,\theta}_{\rho\mu}$
 \begin{eqnarray}
 T^{s,\theta}_{\rho\mu} &=& -g_{\rho\mu}{\cal L}^\theta_{inv}
 -F_{\mu\nu}F_\rho^{\ \nu}(1-{1\over2}\theta^{\alpha\beta}F_{\alpha\beta}) 
 +{1\over4}F_{\mu\nu}\theta_\rho^{\ \nu}F^2\nonumber \\
 &&+F_{\mu\nu}\theta_{\rho\beta}F_\alpha^{\ \beta}F^{\nu\alpha}
 -(F_{\mu\alpha}F_{\rho\nu}+F_{\rho\alpha}F_{\mu\nu})F^{\alpha}_{\ \beta}
 \theta^{\nu\beta}.
 \end{eqnarray}
The latter equation confirms Kruglov's result \cite{Kruglov}.
One has to stress 
that $T^{s,\theta}_{\rho\mu}$ is $not$ symmetric and $not$ traceless. 
Additionally, it is remarkable that in expanding the tensor
(\ref{curs}) for the $U(1)$-case one gets
 \begin{eqnarray}
 \hat T_{\rho\mu}^s \arrowvert_{{\cal O}(\theta)} &=&
 -g_{\rho\mu}{\cal L}^\theta_{inv}
 -F_{\mu\nu}F_\rho^{\ \nu}(1-{1\over2}\theta^{\alpha\beta}F_{\alpha\beta})
 \nonumber\\ 
 &&-(F_{\mu\alpha}F_{\rho\nu}+F_{\rho\alpha}F_{\mu\nu})F^{\alpha}_{\ \beta}
 \theta^{\nu\beta}
 +\theta^{\alpha\beta}\partial_\beta(A_\alpha F_{\rho\nu}F_{\mu}^{\ \nu})
 \nonumber \\
 &\neq& T^{s,\theta}_{\rho\mu} 
 \end{eqnarray}
We observe that (ignoring the total derivative) the nonsymmetric 
parts of $T^{s,\theta}_{\rho\mu}$ do not appear in the expansion
of $\hat T^s_{\rho\mu}$. Moreover, these are exactly the terms where 
$\theta_\rho^{\ \nu}$ carries a free index $\rho$. 
For $T^{c,\theta}_{\rho\mu}$ and
$\hat T^c_{\rho\mu}$ we get the analogous result.\\ 
Thus we find that the calculation of the energy-momentum tensor does
not commute with the Seiberg-Witten expansion of fields and Moyal products.

\section{Conclusion}

For noncommutative gauge field models we have studied (at the classical
level) the construction of the various energy-momentum tensors in order
to describe translation invariance of different noncommutative gauge field
theories. Due to the presence of the deformation parameter $\theta^{\mu\nu}$
(as a constant, antisymmetric, fixed tensor) Lorentz and dilation invariances
are manifestly broken, entailing that the corresponding stress tensors are
not symmetric and not traceless. The obtained results may be the basis for 
the discussion of broken Lorentz and dilation symmetry.


\begin{thebibliography}{99}

\bibitem{Gerhold}
A.~Gerhold, J.~Grimstrup, H.~Grosse, L.~Popp, M.~Schweda and R.~Wulkenhaar,
``The energy-momentum tensor on noncommutative spaces: 
Some pedagogical  comments,''
arXiv:hep-th/0012112.

\bibitem{Filk}
T.~Filk,
``Divergencies in a field theory on quantum space,''
Phys.\ Lett.\ B {\bf 376} (1996) 53.

\bibitem{Iorio}
A.~Iorio and T.~Sykora,
``On the space-time symmetries of non-commutative gauge theories,''
arXiv:hep-th/0111049.

\bibitem{Carroll}
S.~M.~Carroll, J.~A.~Harvey, V.~A.~Kostelecky, C.~D.~Lane and T.~Okamoto,
``Noncommutative field theory and Lorentz violation,''
Phys.\ Rev.\ Lett.\  {\bf 87}, 141601 (2001)
[arXiv:hep-th/0105082].

\bibitem{Carlson}
C.~E.~Carlson, C.~D.~Carone and R.~F.~Lebed,
``Bounding noncommutative QCD,''
Phys.\ Lett.\ B {\bf 518} (2001) 201
[arXiv:hep-ph/0107291].

\bibitem{Boresch}
A.~Boresch, S.~Emery, O.~Moritsch, M.~Schweda, T.~Sommer and H.~Zerrouki,
``Application of Noncovariant Gauges in the Algebraic Renormalization
Procedure'', World Scientific.

\bibitem{Callan}
C.~G.~Callan, S.~R.~Coleman and R.~Jackiw,
``A New Improved Energy - Momentum Tensor,''
Annals Phys.\  {\bf 59} (1970) 42.

\bibitem{Coleman}
S.~R.~Coleman and R.~Jackiw,
``Why Dilatation Generators Do Not Generate Dilatations?,''
Annals Phys.\  {\bf 67} (1971) 552.

\bibitem{Abou-Zeid}
M.~Abou-Zeid and H.~Dorn,
``Comments on the energy-momentum tensor in non-commutative field  theories,''
Phys.\ Lett.\ B {\bf 514} (2001) 183
[arXiv:hep-th/0104244].

\bibitem{Jackiw}
R.~Jackiw, ``Gauge-Covariant Conformal Transformations'',
Phys.\ Rev.\ lett.\ {\bf 41}, 1635 (1978)

\bibitem{Bichl}
A.~A.~Bichl, J.~M.~Grimstrup, H.~Grosse, E.~Kraus, L.~Popp, M.~Schweda and
R.~Wulkenhaar,
``Noncommutative Lorentz symmetry and the origin of the Seiberg-Witten  map,''
arXiv:hep-th/0108045.

\bibitem{Kraus}
E.~Kraus and K.~Sibold,
``Conformal symmetry breaking and the energy momentum tensor in 
four-dimensions,'' 
Nucl.\ Phys.\ B {\bf 372} (1992) 113.

\bibitem{Seiberg}
N.~Seiberg and E.~Witten,
``String theory and noncommutative geometry,''
JHEP {\bf 9909} (1999) 032
[arXiv:hep-th/9908142].

\bibitem{Kruglov}
S.~I.~Kruglov,
``Maxwell's theory on non-commutative spaces and quaternions,''
Annales de la Fondation
Louis de Broglie, Vol. 27 (2002) 343-358
[arXiv:hep-th/0110059].

\bibitem{Jackson}
J.~D.~Jackson, ``klassische Elektrodynamik'', Walter de Gruyter - Berlin
- New York 1981

\bibitem{Madore}
J.~Madore, S.~Schraml, P.~Schupp and J.~Wess,
``Gauge theory on noncommutative spaces,''
Eur.\ Phys.\ J.\ C {\bf 16} (2000) 161
[arXiv:hep-th/0001203].

\bibitem{Bichl2}
A.~A.~Bichl, J.~M.~Grimstrup, L.~Popp, M.~Schweda and R.~Wulkenhaar,
``Perturbative analysis of the Seiberg-Witten map,''
arXiv:hep-th/0102044.

\bibitem{Jurco}
B.~Jurco, L.~Moller, S.~Schraml, P.~Schupp and J.~Wess,
``Construction of non-Abelian gauge theories on noncommutative spaces,''
Eur.\ Phys.\ J.\ C {\bf 21} (2001) 383
[arXiv:hep-th/0104153].


\end{thebibliography}
\end{document}